%% file: double_0th.tex
\documentclass{article}
\usepackage{spconf}
\usepackage{graphicx}
\usepackage{amsmath}
\usepackage{mathtools}
\usepackage{epsfig}
\usepackage{float}
\usepackage{lipsum}
\usepackage{stfloats}
\usepackage{array}
\usepackage{amssymb}
\usepackage{cite}
\usepackage{url}
\usepackage{algorithm2e}
\usepackage{algorithmic}
\usepackage{multirow}
\usepackage{fancyh}
\usepackage{upgreek}
\usepackage{dsfont}
\usepackage{gensymb}

\usepackage{multicol, blindtext}
\usepackage{mathtools, cuted}

\usepackage{color}
\usepackage{bm}
\usepackage{booktabs}
\usepackage{tablefootnote}
\usepackage{threeparttable}

\usepackage{multicol, blindtext}
\usepackage{mathtools, cuted}

\usepackage{color}
\usepackage[dvipsnames]{xcolor}
\usepackage{bm}
\usepackage{booktabs}
\usepackage{tablefootnote}
\usepackage{threeparttable}
\usepackage{stfloats}

\DeclareMathOperator*{\argmin}{argmin} % no space, limits underneath in displays

\title{Multi-Rate Variable-Length CSI Compression for FDD Massive MIMO}
\name{Bumsu Park, Heedong Do, Namyoon Lee}
\address{Korea University, Seoul, South Korea}
\input{input.tex}

\begin{document}
%\ninept
%
\maketitle
\begin{abstract}

For frequency-division-duplexing (FDD) systems, channel state information (CSI) should be fed back from the user terminal to the base station. This feedback overhead becomes problematic as the number of antennas grows. To alleviate this issue, we propose a flexible CSI compression method using variational autoencoder (VAE) with an entropy bottleneck structure, which can support multi-rate and variable-length operation. Numerical study confirms that the proposed method outperforms the existing CSI compression techniques in terms of normalized mean squared error.

\end{abstract}
\begin{keywords}
FDD massive MIMO, CSI compression
\end{keywords}
\section{Introduction}
\label{sec:intro}

% 책이 ref하는 paper를 ref except overall proof
% mu mimo에서 csi feedback에 왜 필요한가
% csi feedback -> dof eff up, but at fdd, downlink training, quantization(codebook) send back amount is huge practical fdd 에서는 사용이 어렵다.
% 다양한 approach로 
\color{black}
    The capacity of MIMO channel is contingent on the quality of channel state information (CSI). Without CSI, spatial degree of freedom of downlink channel is same with MISO channel\cite{heath2018foundations}. In general, CSIT is obtained via train at receiver side and feedback to base station (BS). In time domain duplexing (TDD) system, BS can obtain downlink CSI with uplink training CSI thanks to channel reciprocity. However, as frequency domain duplexing (FDD) system use different frequency bands for uplink and downlink, CSI feedback is essential. 
    
    To maintain the feedback quality, the feedback overhead is proportional to the number of transmit antennas. In this regard, for massive MIMO that using hundreds, thousands of antennas, TDD system is favored over FDD system. Having said that, currently deployed LTE networks are dominantly FDD. Therefore, enabling FDD massive MIMO is an important problem.
    
\begin{figure}
    \centering
    \includegraphics[width = 0.9\linewidth]{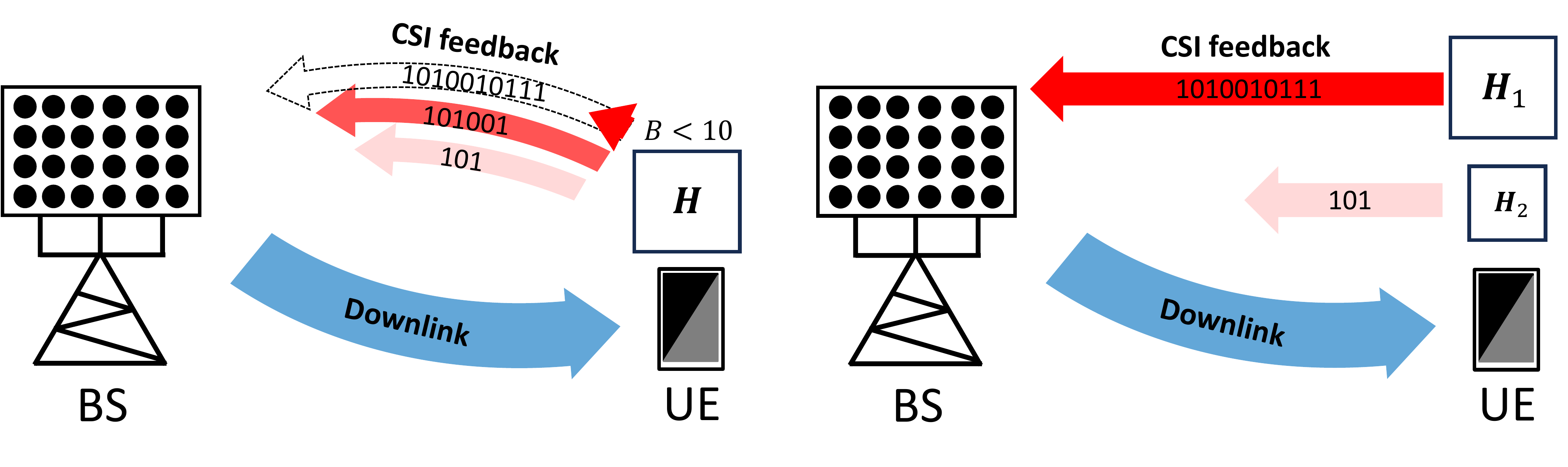}
    \caption{Multi-rate and variable-length operation.}
    \label{fig:systemModel}
\end{figure}

    To bring FDD massive MIMO in reality, significant progress has been made in recent years to reduce CSI feedback overhead\footnote{There is a relevant line of research which targets reducing pilot overhead by exploiting the correlation between the uplink and downlink channels \cite{adhikary2013joint}. A potentially more disruptive direction is entire elimination of downlink channel training by exploiting the channel reciprocity in angle-delay domain \cite{vasisht2016eliminating, han2023fdd}. In both cases, there is little to no need of CSI feedback. Consideration of such possibilities is out of the scope of this paper.} using compressive sensing \cite{gao2015structured} and deep learning \cite{wen2018deep}. Particularly, autoencoder is well-suited for compression tasks as it reduces the dimensionality of the input. The exploding number of papers evidences the increasing interest on the matter \cite{guo2022overview}. However, most existing deep learning based methods have one of the following drawbacks:
\begin{itemize}
    \item As pointed out in \cite{ornhag2023critical}, they are oblivious in bit level compression. They only consider dimensionality reduction of input. 
    \item They only support single compression rate (see Fig. \ref{fig:systemModel}). With one model support serving a single operating point, the amount of memory required to stored models is proportional to the number of operating points, which is impractical espacially for mobile devices. 
    \item They output fixed-length codewords (see Fig. \ref{fig:systemModel}). Recent studies have not provided a criterion for how many bits to allocate for a given CSI. The deep learning based methods have focused on maximizing only the average performance in a given dataset. As pointed out in \cite{kim2022learning}, enormous imbalance of distortion occurs due to uniform bit allocation.
\end{itemize}

There are a handful of studies addressing two out of the three issues. The methods in \cite{liang2022changeable, nerini2022machine, ornhag2023critical} and \cite{yang2019deep, ravula2021deep, kim2022learning} supports multi-rate and variable-length operation, respectively, and all of them are quantization-aware. 
The present paper proposes a deep learning based CSI compression architecture, which
\begin{itemize}
    \item supports multi-rate and variable-length operation while being quantization-aware.
    \item exhibits a better trade-off in terms of normalized mean squared error (NMSE) and the number of bits.
\end{itemize}

% \begin{figure}
%     \centering
%     \includegraphics[width = 0.99\linewidth]{example-image}
%     \caption{System model.}
%     \label{fig:systemModel}
% \end{figure}
\hfill
\section{System Model}
\color{black}
Consider a downlink channel training for FDD massive MIMO channel. The BS with $N_t$ numbers of antenna with linear antenna array sends a pilot signal to the single-antenna user terminal. We consider the system employing orthogonal frequency division multiplexing(OFDM) with $N_c$ subcarriers. In alignment with other prior studies\cite{liang2022changeable, nerini2022machine, ornhag2023critical, yang2019deep, ravula2021deep, kim2022learning}, we assume that user can reconstruct perfect channel, which is not in practice. 
The downlink channel can be compactly represented by $N_c\times N_t$ size matrix $\boldsymbol{H}$. Although the physical channel are in a high dimensional space, it can be represented with only a few variables, therefore they can be efficiently described with a small number of parameters\cite{he2021wireless}.

\subsection{Preprocessing}
\color{black}
The angular-delay domain representation of the channel can be written as $\boldsymbol{F}_{N_c}\boldsymbol{H}\boldsymbol{F}_{N_t}^{H}$ where $\boldsymbol{F}_{N}$ is the $N$-point discrete Fourier transform matrix, that is, $[\boldsymbol{F}_N]_{m,n}=e^{-j2\pi{{(m-1)\times (n-1)} \over N}}$. Since large delay brings large path loss, high index components in row of $\boldsymbol{F}_{N_c}\boldsymbol{H}\boldsymbol{F}_{N_t}^{H}$ are almost zero. In this regard, following the convention proposed in \cite{wen2018deep}, we hereafter work with the preprocessed channel
\begin{align}
    \bar{\bH} = f_{\sf trun}(\bF_{\Nc}\bH\bF_{N_t}^H)
\end{align}
with $f_{\sf trun}$ is an operation that keeps first $\bar{N}$ columns and truncate the remainders. 

\subsection{Problem statement}
\color{black}
The problem is, with a given bit budget $B\in \mathbb{N}$, to compress the channel into a bit stream and decompress it with smallest distortion as possible. In information-theoretical jargon, our goal is to design a encoder and decoder 
\begin{align}
    f_{\sf enc}:\mathbb{C}^{\bar{N_c}\times N_t} \rightarrow \{0,1\}^+
\,\,\,\,
    f_{\sf dec}:\{0,1\}^+ \rightarrow \mathbb{C}^{\bar{N_c}\times N_t}
\end{align}
that minimizes distortion under bit budget constraint
\begin{align}
    D &= \mathbb E_{\boldsymbol{\bar H}}[d(f_{\sf dec}(f_{\sf enc}(\boldsymbol{\bar H})),\boldsymbol{\bar H})]\\
    R &= \ell(f_{\sf enc}(\boldsymbol{\bar H})) \le B \nonumber
\end{align}
where $\{0,1\}^+$ is a set of all finite-length binary strings and $\ell(\cdot)$ is the length of string. The distortion function $d$ is set by NMSE in this work. That is,
\begin{align}
    d(f_{\sf dec}(f_{\sf enc}(\boldsymbol{\bar H})),\boldsymbol{\bar H}) = {{\Vert {d(f_{\sf dec}(f_{\sf enc}(\boldsymbol{\bar H}))-\boldsymbol{\bar H})}\Vert}^{2}_{\sf F}\over{\Vert \boldsymbol{\bar H}\Vert}^{2}_{\sf F}} 
\end{align}
where $\Vert \cdot \Vert^{2}_{\sf F}$ denotes the Frobenius norm.

\section{Proposed Method}
\color{black}
\subsection{Entropy bottleneck}
\begin{figure}
    \centering
    \includegraphics[width = 0.8\linewidth]{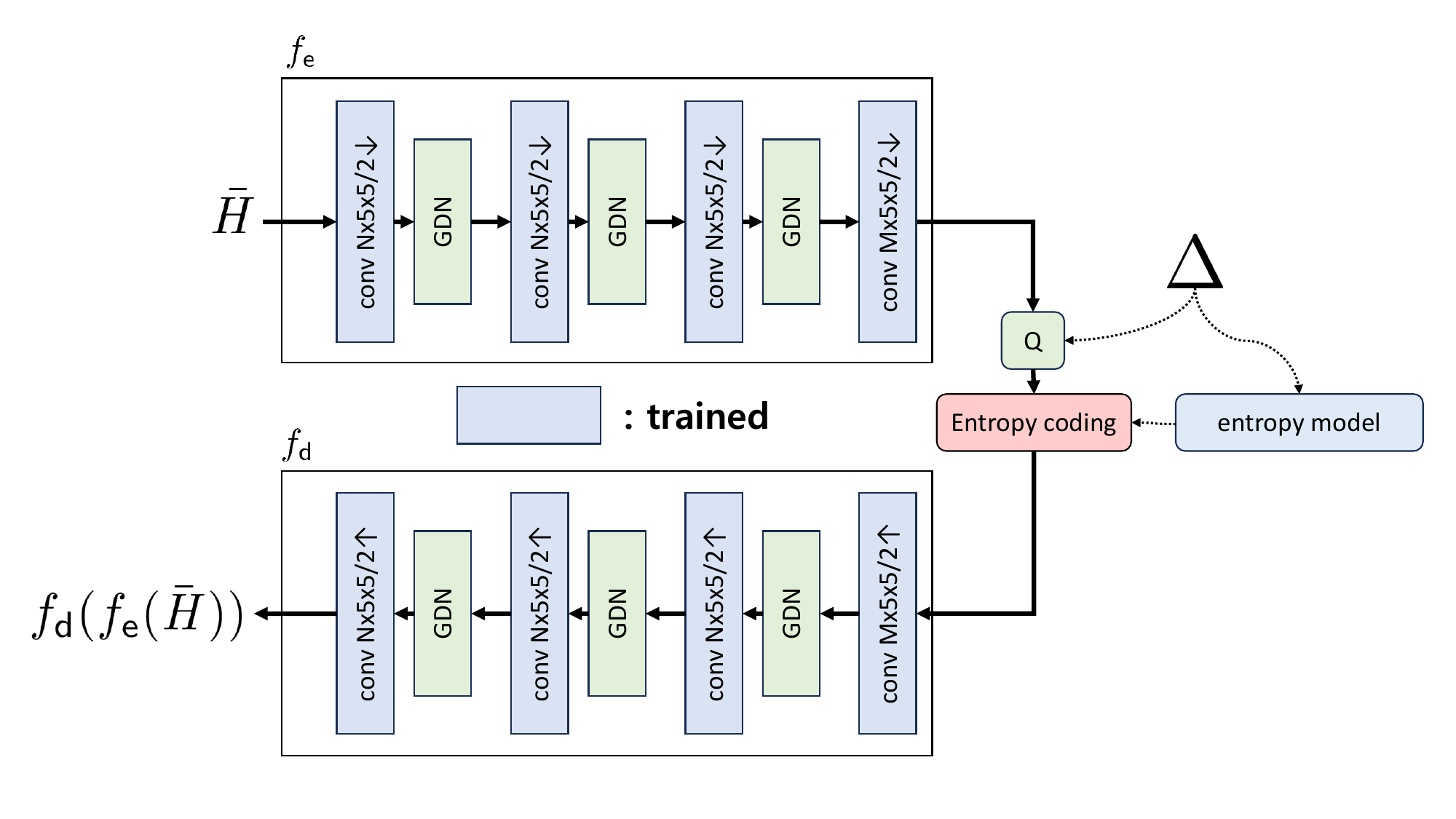}
    \caption{Proposed CSI compression architecture.}
    \label{fig:architecture}
\end{figure}

We adopt the entropy bottleneck architecture in \cite{balle2018variational} (see Fig. \ref{fig:architecture}). Entropy bottleneck is an architecture for end-to-end training image codec, which is quantization-aware and supports variable-length. As in Fig. \ref{fig:architecture}, entropy bottleneck is composed with encoder $f_{\sf e}(\,\cdot\,;\boldsymbol{\theta}_{\sf e})$, decoder $f_{\sf d}(\,\cdot\,;\boldsymbol{\theta}_{\sf d})$ and entropy model $p_{\boldsymbol{y}}(\,\cdot\,;\psi)$. Encoder-decoder is a pair of transform functions that maps between spaces of image $\boldsymbol{x}$ and continuous latent $\boldsymbol{y}=f_{\sf e}(\,\boldsymbol{x}\,;\boldsymbol{\theta}_{\sf e})$. The goal of entropy bottleneck is finding the best possible parameter set $\boldsymbol{\theta}_{\sf e}, \boldsymbol{\theta}_{\sf d}$ that minimize the loss function 
\begin{align}
    \label{lossfunction}
    \min_{{\btheta}_{\sf e},\btheta_{\sf d}}\, &D + \lambda R\\
    \text{where}\, &D = \mathbb{E}_{\boldsymbol{x}} [d(\boldsymbol{x},f_{\sf d}(\boldsymbol{\hat y}\,;\boldsymbol{\theta}_{\sf d})]\nonumber\\ 
    &R = -\sum p_{\boldsymbol{\hat y}}\log p_{\boldsymbol{\hat y}}\nonumber
\end{align}
for $\boldsymbol{{\hat y}} \coloneqq Q(\boldsymbol{y};\Delta)$. Quantization function Q is parameterized by the bin size $\Delta$ and the maximum quantization level $l_{\sf max}$. Precisely, it applies
\begin{align}
    y \mapsto \argmin_{\hat y \in \Delta\cdot(\bbZ\cap [-l_{\sf max},l_{\sf max}])} |\hat y-y|
    \label{quantization}
\end{align}
in a component-wise manner. Lastly, the trainable parameter $\boldsymbol{\psi}$ is required to construct a differentiable proxy of the probability distribution function (PDF) of encoder output of each input.

\color{black}
\subsection{Training}

During training time, the rate and distortion term in loss function \eqref{lossfunction} is not differentiable due to quantization function. To handle this problem, authors in \cite{balle2018variational} adopted uniform noise
\begin{align}
    {\hat y}\approx {\tilde{y}} = {y} + {\sf Unif}[-{\Delta / 2},{\Delta / 2}] 
    \label{latentapprox}
\end{align}
in a component-wise manner. This mimics the effect of quantization and make the loss function \eqref{lossfunction} differentiable. With factorized latent assumption, the PMF of proxy latent $p_{{\tilde y}}$ is calculated as 
\begin{align}
    p_{{\tilde y}}(n) = \int^{n+{\Delta/2}}_{n-\Delta/2} p_{{y}}\,dy
\end{align}
with $\,n\in \Delta\cdot(\bbZ\cap [-l_{\sf max},l_{\sf max}])$. Then, the rate in \eqref{lossfunction} is reformulated as
\begin{align}
    &R = -\sum p_{\boldsymbol{\tilde y}}\log p_{\boldsymbol{\tilde y}}
\end{align}

\subsection{Inference}
In inference time, with trained $p_{\boldsymbol{y}}$ and $Q(\,\cdot\,;\Delta)$, we can perform lossless source coding, i.e. arithmetic coding, with rate of $R$. As we do not estimate $p_{\boldsymbol{\hat y}}$ directly but $p_{\boldsymbol{y}}$, we can make use of \eqref{latentapprox} for any value of $\Delta \in [0,\infty)$ provided that the approximation is accurate. Fixing the trained parameters $\{\boldsymbol{\theta}_{\sf e},\boldsymbol{\theta}_{\sf d},\boldsymbol{\psi}\}$, larger $\Delta$ reduces the entropy $R$, resulting in higher compression rate. This simple trick enables multi-rate operation with a single trained model.

As the value of $\Delta$ should be informed to the decoder, we need to use a finite dictionary $\cD$ instead of $[0,\infty)$. It technically requires additional $\lceil \log_2 |\cD| \rceil$ bits, however we do not take it into account since the number of additional bits is negligible and it makes hard to compare the proposed architecture with existing methods.

\subsubsection{Modification for fixed-length operation}
In general, variable-length architecture perform better than fixed-length architecture since the constraint on bit length is relaxed. Even though, fixed-length architecture is often preferred than variable-length architecture for reasons such as simplicity, standardization. Our multi-rate method can operate fixed-length feedback by simple modification to transmitting bit sequence. First, select the smallest $\Delta$ satisfying
\begin{align}
    \mathbb{E}_{\boldsymbol{\bar {H}}} [\ell(f_{\sf ent}(f_{\sf e}(\boldsymbol{\bar{H}};\boldsymbol{\theta}_{\sf e});\boldsymbol{\psi}))] \le B
\end{align}
where $f_{\sf ent}(\,\cdot\,;\boldsymbol{\psi})$ is an entropy coding operation.
The constraint in the first place is equality, however it can be easily met by zero-padding. At receiver side, the entropy decoder terminate the decoding whenever $K$ symbols are already decoded. Again, the value of $\Delta$ should be informed to the decoder.
\begin{table}[]
\centering
\begin{threeparttable}
\setlength{\tabcolsep}{2pt}
\caption{Baselines, all of which are quantization-aware.}
\label{table:priorArt}
\begin{tabular}{lcccc} 
\toprule
Name & Ref. & Multi-rate & Variable-length  \\
\midrule
Transform coding\tnote{*} & \cite{ornhag2023critical}  & \checkmark &\\
CH-CsiNetPro-PQB & \cite{liang2022changeable} & \checkmark &\\
Ordered vector quantization & \cite{rizzello2023user}  & \checkmark &\\
Entropy bottleneck & \cite{ravula2021deep} & & \checkmark \\
{\bf Proposed} & - & \checkmark & \checkmark \\
\bottomrule
\end{tabular}
\begin{tablenotes} \footnotesize
\item[*] We first pick the top-$K_{\sf TC}$ entries of $\bar{\bH}$ in terms of absolute value, and then allocate $B_{\sf TC}$ bits per dimension. Here, $K_{\sf TC}$ is the largest integer abiding by the bit budget constraint, $2K_{\sf TC}(\log_2 \bar{N}_{\mathrm{c}}M + B_{\sf TC})\leq B$. For each $B$, we choose the best $B_{\sf TC}\in\{2,4,8,16\}$ in terms of NMSE performance.
\end{tablenotes}
\end{threeparttable}
\vspace*{-4mm}
\end{table}

\begin{figure}
    \centering
    \includegraphics[width = 1\linewidth]{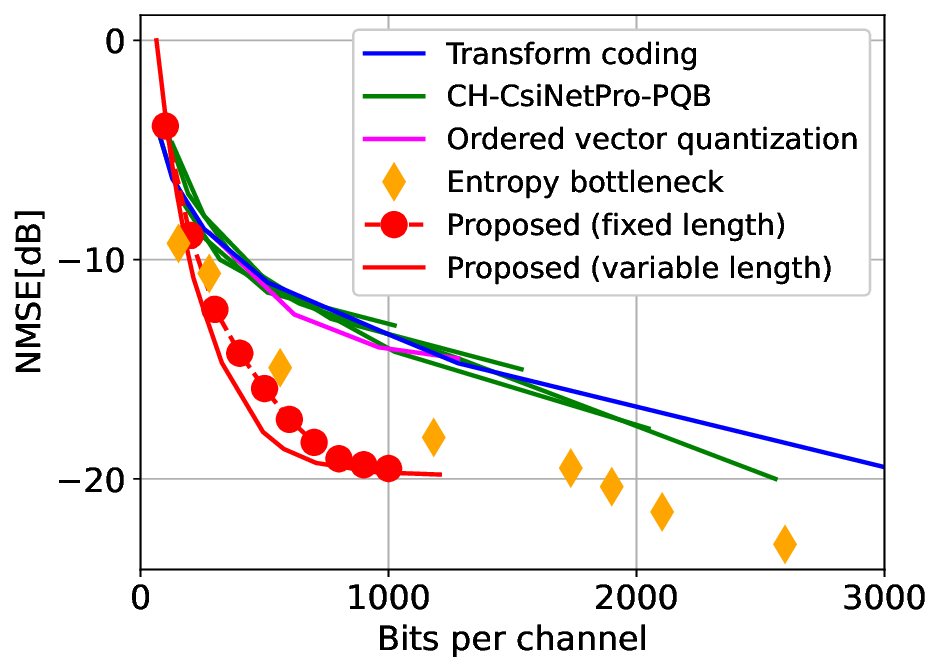}\vspace*{-5mm}
    \caption{NMSE performance with respect to the bit budget.}\vspace*{-5mm}
    \label{fig:experiments}
\end{figure}

\section{Experiments}
\subsection{Dataset}
We train and test the model with COST2100 dataset \cite{wen2018deep}, the most widely used benchmark dataset for CSI compression.
% The data are generated by default setting in COST2100 MIMO channel model \cite{6393523} with $M = 32, \Nc =1024, \bar{N}_{\mathrm{c}}=32$.
Specifically, we consider indoor 5.3GHz scenario with $(M,\Nc,\bar{N}_{\mathrm{c}}) = (32, 1024, 32)$. Number of samples are 100,000, 30,000, 50,000 for train, validation, and test. 

\subsection{Hyperparameters}
For training, we set the hyperparameters to $\lambda=5\cdot 10^{-4}$ and  $\Delta=1$. Note that this setup corresponds to $B=493.69$, which can be measured by averaging $\ell(f_{\sf e}(\bar{\bH}))$. 
% and $\text{NMSE}=-18.58$ dB. 
For inference, we choose $\Delta\in \cD$ for multi-rate operation where 
% \heedong{How about something more ``natural''? Say $\{1.2^{-12}, 1.2^{-11}, \ldots, 1.2^{19}\}$}
\begin{align}
    \cD = \{0.125,0.25,0.5,0.75, 1,2,3,\ldots,16, 32\}.
\end{align}

% All experiments are implemented on colab-pro-plus environment. 

% We use the same dataset as \cite{liang2022changeable} for fair comparison. The data are generated by default setting in COST2100 MIMO channel model \cite{6393523} with $M = 32, \Nc =1024, \bar{N}_{\mathrm{c}}=32$. We consider indoors 5.3GHz scenario. Number of used samples are 100,000, 30,000, 50,000 for train, validation and test. We set the hyperparamter to be $\lambda=5\cdot 10^{-4}$ and set $\Delta=1$. All experiments are implemented on colab-pro-plus environment.

% \subsubsection{Metrics}
% MSE:
% \begin{align}
%     NMSE(\bH,\hat{\bH})=E\Biggl\{ 
%     {{\Vert\bH - \hat{\bH}\Vert^2_2}\over {{\Vert \bH}\Vert}^2_2} 
%     \Biggr\}.
% \end{align}

% \subsection{Baselines}
% \begin{itemize}
%     \item Transform coding \cite{ornhag2023critical}:
%     \item Feedback overhead control unit (FOCU) \cite{liang2022changeable}:
%     \item Ordered vector quantization (OVQ) \cite{rizzello2023user}:
%     \item Entropy bottleneck \cite{ravula2021deep}: 
% \end{itemize}

\subsection{Results}
\color{black}
Fig. \ref{fig:experiments} depicts the NMSE performance of CSI compression methods, listed in Table. \ref{table:priorArt}, with respect to the number of bits used for the feedback. The deep learning based CSI compression methods, all but transform coding, consists of 2.1 million trainable parameters per single trained model. The proposed method, as well as the one with fixed-length modification, outperforms all multi-rate methods by a large margin. In comparison to the entropy bottleneck, ours achieves superior results. This performance gain is attributed to architectural differences, and although further performance optimization through  architectural fine-tuning is possible, it falls beyond the scope of our research. 

\begin{figure}
    \centering
    \includegraphics[width = 1\linewidth]{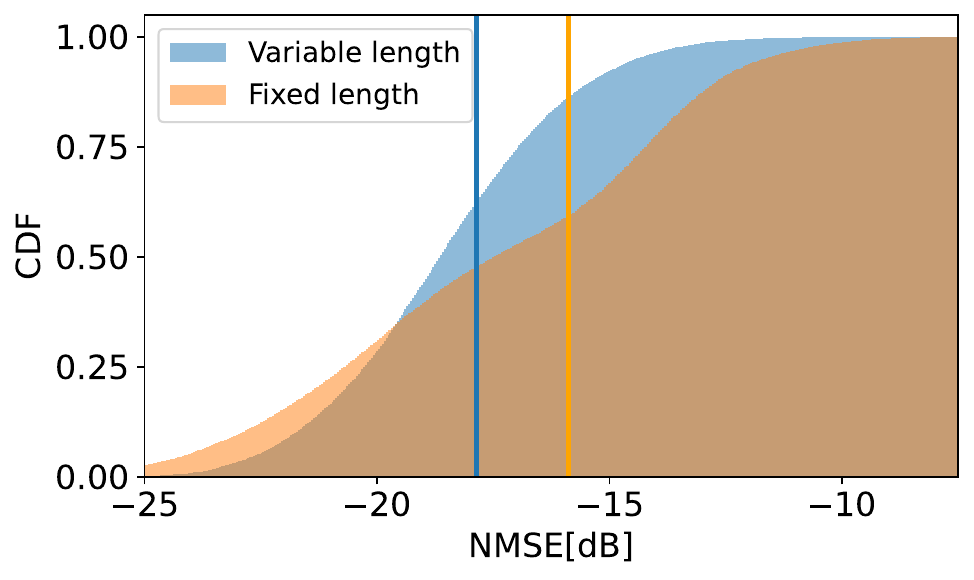}\vspace*{-4mm}
    \caption{CDF of proposed methods with $B=500$. Each vertical line corresponds to the average NMSE of each method.}\vspace*{-3mm}
    \label{fig:variableAdvantage}
\end{figure}

Fig. \ref{fig:variableAdvantage} also shows the advantage of our variable-length CSI compression method over the fixed-length counterpart. As can be seen, the variable-length CSI compression method outperforms the fixed-length method.
% in terms of both average distortion and variance.
% In Fig. \ref{fig:experiments}, we show the NMSE between $\bar{\bH}$ and $f_{\sf d}(f_{\sf e}(\bar \bH))$ with respect to the number of bits used for the feedback. As classified in Table. \ref{table:priorArt}, we compare our method with multi-rate CSI compression methods and also with Entropy Bottleneck \cite{ravula2021deep}, which employs a structure similar to ours but does not consider multi-rate scenarios. The deep learning based CSI compression methods, except the DCT-based transform coding, consists of 2.1 million trainable parameters per single trained model, which corresponds to a single line or an isolated marker in Fig. \ref{fig:experiments}. Shown in Fig. \ref{fig:experiments}, our method outperforms other multi-rate methods by a large margin. It is worth noting that, even with variable-length and fixed-length modifications imposing equality constraints on bit budgets, our proposed scheme maintains superior performance among multi-rate methods. In comparison to the entropy bottleneck with same number of parameters, ours achieves superior results. This performance gain is attributed to architectural differences, and although further performance optimization through  architectural fine-tuning is possible, it falls beyond the scope of our research. 

% In this investigation, we undertake a comprehensive comparison of baseline methodologies. Table 1 provides a detailed overview of the methods employed. We evaluate Multi-rate CSI compression techniques and also include a performance assessment of Entropy Bottleneck [18], which shares a similar architecture with ours but does not consider multi-rate scenarios. Our performance evaluation is based on the Normalized Mean Square Error (NMSE) concerning feedback bits per channel, as illustrated in Figure 3. With the exception of the DCT-based transform coding, the remaining methods are rooted in deep learning-based CSI compression approaches. Each method is characterized by a distinct parameter count: CH-CsiNetPro-PQB, Ordered vector quantization, and our proposed method utilize 2.1 million parameters per line, while Entropy Bottleneck employs 2.1 million parameters per point. As evident in Figure 3, our proposed scheme consistently outperforms other multi-rate CSI feedback methods across various domains. 

% \heedong{Things worth mentioning: The data points for EB is extracted from \cite[Fig. 4]{ravula2021deep} using a tool called {\fontfamily{qcr}\selectfont webPlotDigitizer}. Details for transform coding. 
% }

\section{Conclusion}
\color{black}
% We introduce a new framework for multi-rate CSI compression, which is essential for practical CSI feedback. The main innovation involves using the learned CDF to calculate the PMF of quantized values when different quantization bin sizes are applied, which allows us to utilize a single model trained at a single rate for multi-rate CSI compression. Our framework, with just a simple tweak, enables a single model to perform multi-rate CSI compression over a wide bit range and  demonstrates superiority over existing multi-rate CSI compression methods in wide bit range.
We introduce a new framework for multi-rate variable length CSI compression, which is essential for practical CSI feedback. The key idea is to quantize with different operating bin size, using estimated latent CDF. Then, BS can receive compressed CSI with cost of finite bits. This simple idea enables a single model to perform multi-rate CSI compression over a wide bit range. Notably, the proposed model even exhibits a better NMSE performance than an ensemble of single-rate models.

\bibliographystyle{IEEEbib}
\bibliography{ref}

\end{document}

%% file: input.tex
\def\bb0{{\mathbb{0}}}

\def\bb{{\boldsymbol{b}}}

\def\b0{{\boldsymbol{0}}}

\def\bF{{\boldsymbol{F}}}

\def\bH{{\boldsymbol{H}}}

\def\b{{\mathrm{b}}}

\def\r0{{\mathbf{0}}}

\def\bbZ{{\mathbb{Z}}}

\def\cD{\mathcal{D}}

\def\btheta{\bm \theta}

\def\bsf0{{\bm{\mathsf{0}}}}

\def\Nc{{N_{\mathrm{c}}}}

\def\N0{{N_{\mathrm{0}}}}

\def\bsf{{\boldsymbol{s}_\mathrm{f}}}

\newcommand{\be}{\begin{equation}}
\newcommand{\ee}{\end{equation}}
\newcommand{\bal}{\begin{align}}
\newcommand{\eal}{\end{align}}